\documentclass[acmsmall,screen]{acmart}

\AtBeginDocument{%
  \providecommand\BibTeX{{%
    Bib\TeX}}}

\usepackage[utf8]{inputenc}
\usepackage[T1]{fontenc}
\usepackage{amsmath,amsfonts} 
\usepackage{graphicx}
\usepackage{textcomp}
\usepackage{xcolor}
\usepackage{url}
\usepackage{pifont}
\usepackage{tabularx}
\usepackage{makecell}
\usepackage{multirow}
\usepackage{xspace}

\newcommand{\tool}{\textsc{Flare}\xspace}

\definecolor{todocolor}{rgb}{0.9,0.1,0.1}
\definecolor{donecolor}{rgb}{0.1,0.6,0.4}

\newcommand{\eg}{\hbox{\emph{e.g.}}\xspace}
\newcommand{\ie}{\hbox{\emph{i.e.}}\xspace}

\usepackage{pythonhighlight}

\definecolor{codegreen}{rgb}{0,0.6,0}
\definecolor{codegray}{rgb}{0.5,0.5,0.5}
\definecolor{codepurple}{rgb}{0.58,0,0.82}
\definecolor{backcolour}{rgb}{0.97,0.97,0.95}
\definecolor{forestgreen}{rgb}{0.28,0.62,0.37}

\lstdefinestyle{mystyle}{
    backgroundcolor=\color{backcolour},   
    commentstyle=\color{codegray},
    keywordstyle=\color{codepurple},
    numberstyle=\tiny\color{codegray},
    stringstyle=\color{blue},
    basicstyle=\ttfamily\footnotesize,
    breakatwhitespace=false,         
    breaklines=true,                 
    captionpos=b,                    
    keepspaces=true,                 
    numbers=left,                    
    numbersep=5pt,                  
    showspaces=false,                
    showstringspaces=false,
    showtabs=false,                  
    tabsize=4,
}

\lstdefinestyle{SQL}{
    language = SQL,
    backgroundcolor=\color{backcolour},     
    basicstyle = \small\ttfamily,           
    rulesepcolor= \color{gray},             
    breaklines = true,                  
    numbers = left,                     
    numbersep=5pt,                  
    showspaces=false,                
    showstringspaces=false,
    showtabs=false,                  
    tabsize=4,
    numberstyle = \tiny\color{codegray},               
    keywordstyle = \color{blue},            
    commentstyle =\color{codegray},        
    showspaces = false,                 
}

\setcopyright{none}             
\makeatletter
\renewcommand\@formatdoi[1]{\ignorespaces} 
\makeatother
\def\BibTeX{{\rm B\kern-.05em{\sc i\kern-.025em b}\kern-.08em
    T\kern-.1667em\lower.7ex\hbox{E}\kern-.125emX}}

\usepackage{algorithm}
\usepackage{algpseudocode}
\usepackage{threeparttable}

\begin{document}

\title{FLARE: Agentic Coverage-Guided Fuzzing for LLM-Based Multi-Agent Systems}

\author{Mingxuan Hui}
\affiliation{%
  \institution{Xidian University}
  \city{ShaanXi}
  \country{China}}
\email{25031111082@stu.xidian.edu.cn}

\author{Xinyue Li}
\affiliation{%
  \institution{Peking University}
  \city{Beijing}
  \country{China}}
\email{xinyueli@stu.pku.edu.cn}

\author{Lu Wang}
\affiliation{%
  \institution{Xidian University}
  \city{ShaanXi}
  \country{China}}
\email{wanglu@xidian.edu.cn}

\author{Chengcheng Wan}
\affiliation{%
  \institution{East China Normal University}
  \city{Shanghai}
  \country{China}}
\affiliation{%
  \institution{Shanghai Innovation Institute}
  \city{Shanghai}
  \country{China}}
\email{ccwan@sei.ecnu.edu.cn}

\author{Yifan Wang}
\affiliation{%
  \institution{Xidian University}
  \city{ShaanXi}
  \country{China}}
\email{22009200272@stu.xidian.edu.cn}

\author{Yimian Wang}
\affiliation{%
  \institution{Xidian University}
  \city{ShaanXi}
  \country{China}}
\email{23009200555@stu.xidian.edu.cn}

\author{Feiyue Song}
\affiliation{%
  \institution{Xidian University}
  \city{ShaanXi}
  \country{China}}
\email{25031212122@stu.xidian.edu.cn}

\author{Beining Shi}
\affiliation{%
  \institution{Xidian University}
  \city{ShaanXi}
  \country{China}}
\email{benin@stu.xidian.edu.cn}

\author{Yixi Li}
\affiliation{%
  \institution{Xidian University}
  \city{ShaanXi}
  \country{China}}
\email{23009201088@stu.xidian.edu.cn}

\author{Yaxiao Li}
\affiliation{%
  \institution{Xidian University}
  \city{ShaanXi}
  \country{China}}
\email{yx_li@stu.xidian.edu.cn}

\begin{abstract}
Multi-Agent LLM Systems (MAS) have been adopted to automate complex human workflows by breaking down tasks into subtasks. However, due to the non-deterministic behavior of LLM agents and the intricate interactions between agents, MAS applications frequently encounter failures, including infinite loops and failed tool invocations. Traditional software testing techniques are ineffective in detecting such failures due to the lack of LLM agent specification, the large behavioral space of MAS, and semantic-based correctness judgment.

This paper presents \tool, a novel testing framework tailored for MAS. \tool takes the source code of MAS as input and extracts specifications and behavioral spaces from agent definitions. Based on these specifications, \tool builds test oracles and conducts coverage-guided fuzzing to expose failures. It then analyzes execution logs to judge whether each test has passed and generates failure reports. Our evaluation on 16 diverse open-source applications demonstrates that \tool achieves 96.9\% inter-agent coverage and 91.1\% intra-agent coverage, outperforming baselines by 9.5\% and 1.0\%. \tool also uncovers 56 previously unknown failures unique to MAS.
\end{abstract}


\maketitle

\section{Introduction}\label{sec:intro}

\subsection{Motivation}

LLM-based Multi-Agent Systems (MAS) have emerged as a robust paradigm for replicating and enhancing complex human workflows. By decomposing intricate objectives into granular subtasks, these systems enable specialized agents to collaborate toward a unified goal. However, the non-deterministic nature of Large Language Models (LLMs) introduces stochasticity into agent execution and communication, often resulting in high failure rates. Correctness in MAS relies on the precise coordination of all agents, a guarantee that probabilistic models cannot inherently provide. For instance, agents may inadvertently disclose sensitive information or corrupt transactional data, leading to unintended system behaviors~\cite{DBLP:journals/corr/abs-2412-14470}. Even with state-of-the-art LLMs (\eg, GPT-4.1\cite{openai2025gpt41} and Gemini-3.0-Pro\cite{google2025gemini}), MAS still suffer high failure rates~\cite{pan2025why}. Therefore, how to effectively address these system failures has become an urgent problem. 

\begin{figure}
    \centering
    \includegraphics[scale=0.9]{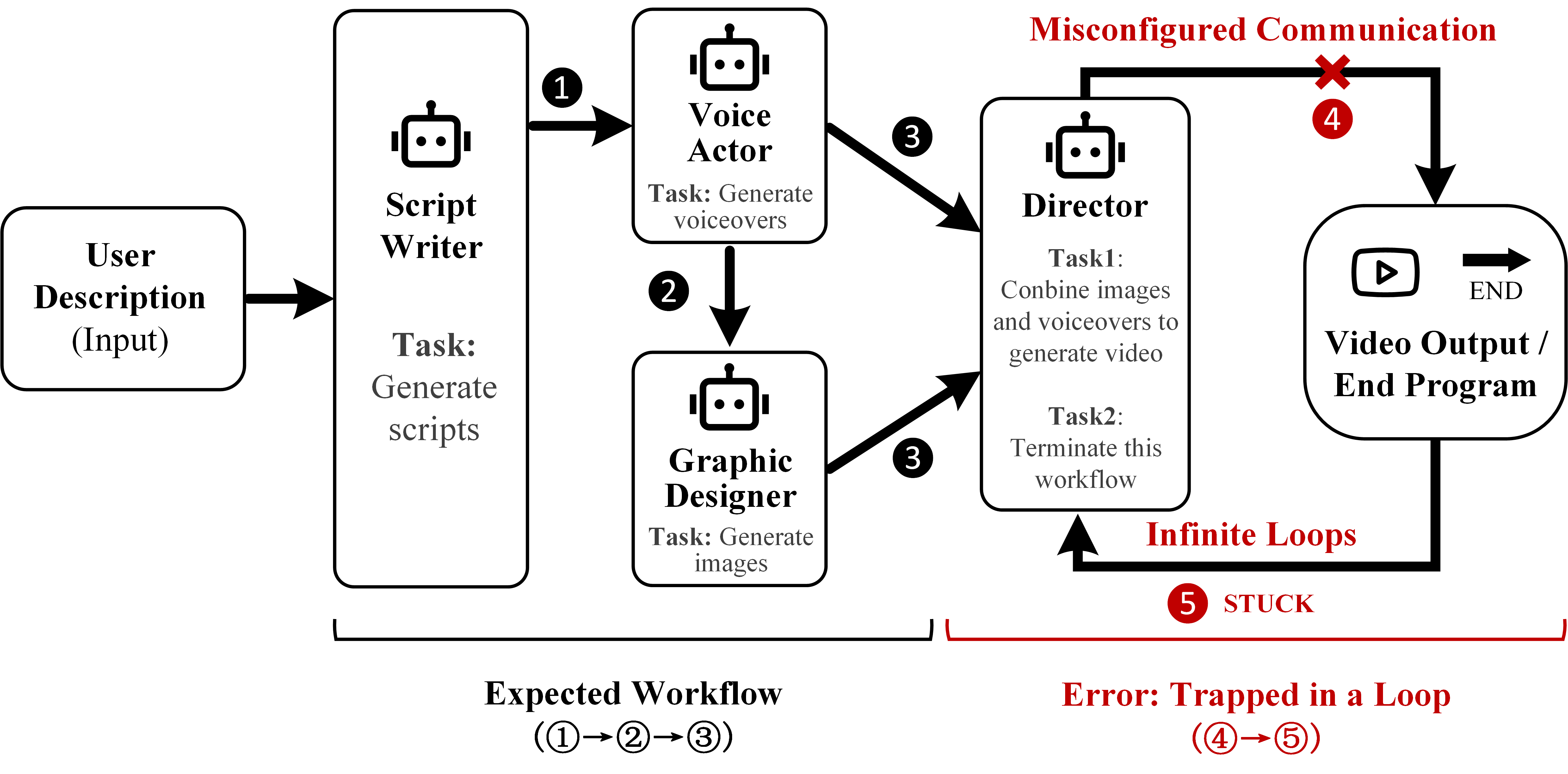}
    \caption{ShortsMaker~\cite{autogen_workflow_gswithjeff} creates videos with four agents.
    }
    \label{fig:1.1}
\end{figure}

To better understand the testing tasks for MAS, we take ShortsMaker~\cite{autogen_workflow_gswithjeff}, a video creation application, as an example. As shown in Figure~\ref{fig:1.1}, given a user description, four agents work collaboratively to generate a video with both voice and graphics. Each agent has its own task, with two data dependencies: (1) the \textit{voice actor} and \textit{graphic designer} rely on the output of the \textit{script writer}; and (2) the director relies on all other agents. While simple, it suffers a severe error (red part), where the director agent is expected to end the workflow but actually trapped in a loop. This error is caused by the wrong configuration of communication mode, which does not allow a single agent to speak twice in a row and thus forbids the director agents to end the workflow after generating the video.

This defect is hard to reveal by traditional testing tools and developers. Traditional testing tools are not able to traverse the huge input space of MAS and lack the capability of judging functionality correctness. Meanwhile, the developers tend to only write 1-2 test cases for the main functionality, which can hardly cover all the possible agent behaviors. 

The above example demonstrates several open challenges in testing MAS:

\medskip
\textbf{1)Lack of LLM agent specifications}. Software testing typically relies on specifications to define both expected and unexpected software behaviors. However, agent specifications, including intended behaviors and interaction rules, are often defined informally in free text that tends to be vague and imprecise. In addition, these definitions are usually dispersed across various software components instead of consolidated in a unified description, including LLM prompts, agent descriptions, code comments, and YAML/JSON configurations. Such fragmentation casts difficulty in obtaining a precise and machine-readable specification for the entire MAS.

Existing methods are ineffective in obtaining agent specifications. 
For non-AI software, traditional code analysis and LLM-based solutions obtain assertion-based software specifications from source code documentation, and describe specifications in mathematical logic expressions~\cite{10.1145/1189748.1189752,hossain2025doc2oracll}. Unfortunately, such assertions are unavailable in MAS. 
Another line of work~\cite{wan2022automated,xie2022towards,zhang2018deeproad} targets AI-enabled software, and propose task specifications for ML tasks with categorical outputs. These specifications are all manually derived from the semantics of the predefined task. 
However, LLM agents have open-ended outputs and support in-context learning, making it impossible to cover all possible tasks through enumeration.

\medskip
\textbf{2)Large behavioral space in Multi-Agent Systems}. Unlike the behavioral space of traditional software that explicitly defined by code structure, the behavioral space of an agent is implicitly defined by the prompt and external tools invoked by LLMs. In addition, the interaction between agents also requires analyzing the communication policy and task dependency to characterize the joint state of the entire MAS. The generative nature of LLMs and higher-level behaviors (\eg, self-planning) also makes it hard to model all the possible actions in a static manner. Therefore, how to obtain the behavioral space of MAS remains an open problem.

To our best knowledge, there is no existing work that formally characterizes the behavioral space of MAS or even LLM. Targeting specific types of software, existing studies define the behavioral space based on the task scenario. Some work proposes domain-specific solutions, including UI states for GUI applications~\cite{neelofar2022instance} and vehicle states~\cite{10064002}. These solutions cannot be applied to MAS.

\medskip
\textbf{3)Semantic-based correctness judgment}. The behavior of MAS is both defined by data-flow and control-flow, of which most failures are non-fatal stop ones. Therefore, judging whether a test passes requires a semantic-based correction judgment that focuses on the functionality and information delivered through tokens. Furthermore, huge logs are generated during MAS execution, recording the behavior and states of different system components, including LLM input/output tokens, agent interactions, external tool invocations, and other agent-unrelated information. Judging correctness from such logs requires content filtering, cross-component validation, and multi-turn conversation analysis.

Prior work makes correctness judgments based on control-flow and data-flow. Control-flow-based solutions~\cite{nikolic2014reachability, soremekun2021locating, wan2022automated} regard a test as passed when the execution path is the same as the expected one.  Data-flow-based approaches~\cite{zhang2017automated,10.1109/TSE.1985.232226,rothermel1997safe} use def-use coverage to characterize the correctness. However, they rely on static control-flow structure and data-dependency relations, which do not reflect the functionality of agents.

\subsection{Contribution}
In this paper, we propose \tool (Fuzzing LLM-Based Multi-Agents for Revealing Errors), the first functional testing framework tailored for MAS.

To address the specification challenge, we formally defined the MAS specification with four core parts: agent relationships, termination patterns, tasks execution, and tool calls. \tool first conducts static analysis to extract MAS-related implementation from source code, and obtains the definitions of intra-agent and inter-agent behaviors. It then invokes a specification agent to transform this information into a structured representation.

To address the behavioral space challenge, we model the behavioral space along two orthogonal dimensions: agent tasks and state transitions. As explicit documentation is often absent, \tool focuses on the internal structure of agents and inter-agent communication patterns defined in code. By augmenting extracted code artifacts with the LLM's inherent domain knowledge, \tool models all the feasible actions and the correct subset. We leverage this behavioral space to calculate \emph{intra-agent behavior coverage} and \emph{inter-agent behavior coverage}, serving as the coverage guidance for the fuzzing module of \tool.

To tackle the judgment challenge, \tool extracts the semantic intent from MAS raw execution data. It first restructures the logs into a sequence of semantic events, capturing execution order and dialogue boundaries while significantly reducing data volume. \tool then utilizes a failure agent and a judge agent to conduct semantic-level analysis and identify defects.

Putting everything together, \tool takes the source code of MAS as input. It first extracts specifications and behavioral spaces through agentic solutions. Based on these specification, \tool builds test oracles and generates the seed set for the fuzzer. A coverage-guided fuzzing loop then generates and executes tests with the aim of covering the behavioral space. To enhance coverage, \tool randomly mutates the user input of MAS, definitions of agent capability, and configurations of agent initial sequence, according to a feedback mechanism that adaptively selects mutators targeting uncovered behaviors by dynamically prioritizing seeds and strategies in revealing them.
Once the time limit is reached, \tool terminates the fuzzer and analyzes execution logs to judge whether each test is passed and generate failure reports.

We evaluate \tool on 16 open-source MAS that encompass diverse task scenarios and functional characteristics. Overall, \tool successfully covers 96.9\% of inter-agent actions and 91.1\% for intra-agent coverage, significantly outperforming baselines, which achieved 87.4\% and 90.1\% respectively. It also achieves an average of 94.2\% statement coverage and 81.7\% branch coverage, 7.2\% and 4.9\% higher than baselines respectively. 
In total, \tool identifies 61 unique MAS-specific failures and 5 crash failures, while baselines only identify 2--6 crashes.


\section{Background}\label{sec2}

The evolution of LLMs has empowered agents with significant autonomy in reasoning, planning, and decision-making. In this study, we refer to LLM-driven entities capable of utilizing memory and tools to satisfy user objectives as agents. To address intricate, multi-step problems, developers increasingly adopt MAS. MAS decomposes high-level objectives into coordinated subtasks executed by specialized agents under the orchestration of an LLM. This paradigm has been extensively explored in recent literature~\cite{hong2024metagpt,ye2025masgpt,DBLP:conf/acl/QianDLLXWC0CCL024}. 

To facilitate the development of these complex systems, a growing array of open-source frameworks, such as Microsoft AutoGen~\cite{autogen2023}, CrewAI~\cite{crewai2025}, Camel~\cite{camel_ai}, and Praison~\cite{PraisonAI}, has emerged. These frameworks offer high-level modular abstractions, enabling developers to easily integrate custom tools, manage agent memory, and define interaction topologies. However, despite the conveniences offered by these frameworks, such applications are prone to high risks of failure. Empirical studies have demonstrated that LLM-driven softwares are particularly susceptible to integration failures, which significantly compromise software functionality, efficiency, and safety~\cite{10.1109/ICSE55347.2025.00204,pan2025why}.

\begin{figure}
    \centering\includegraphics[width=0.8\linewidth]{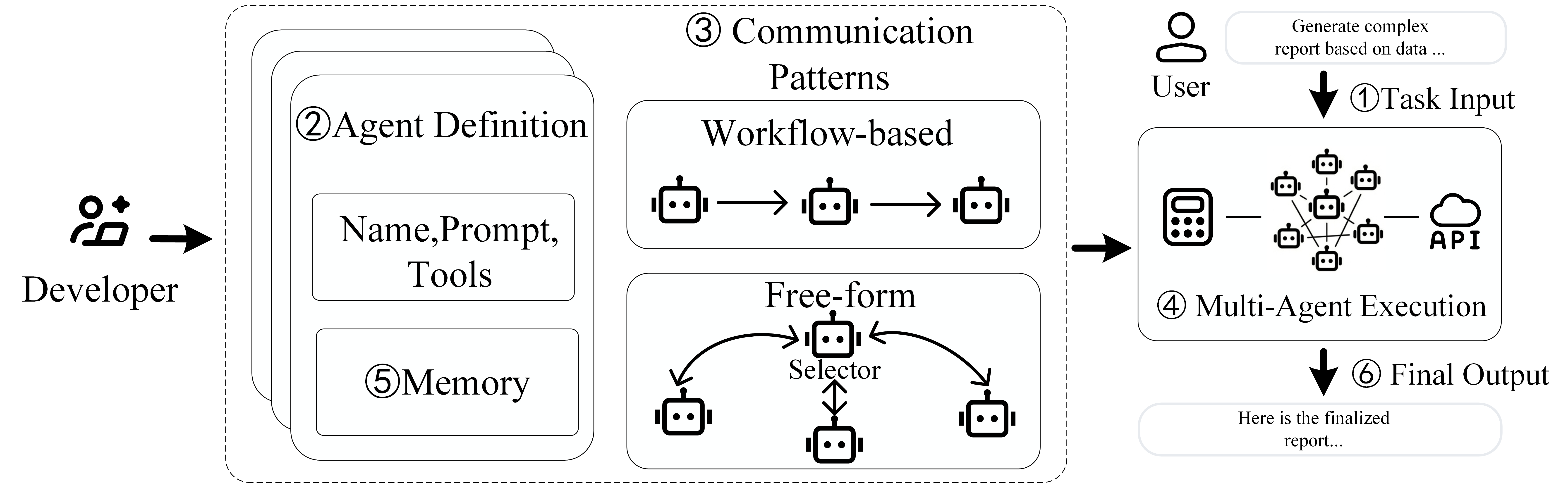}
    \caption{A representative framework of multi-agent systems.}
    \label{2.1}
\end{figure}

Figure~\ref{2.1} illustrates a representative MAS architecture. In this architecture, different agents are assigned specific roles and collaborate according to predefined communication patterns to accomplish task objectives. The system receives tasks through an input port~\ding{172}, where tasks are typically generated by MAS users based on the requirements of a particular application domain and task context.

Subsequently, agents interact according to a communication pattern~\ding{174}. Broadly, communication patterns can be categorized into two classes: workflow-based interaction and free-form interaction. Workflow-based interaction constrains the speaking order of agents via explicit rules; the most straightforward example is to predefine a fixed turn order. This mode affords advantages in controllability but may be limited in adapting to dynamic task contexts. By contrast, free-form interaction allows the speaking order to be determined dynamically based on context, often by a ``selector'' agent that decides the next speaker in each round; this approach requires analyzing and reasoning about latent relations within the agent set (e.g., capability complementarity, dependency structures, and associations with the current working memory). A common example is the selector mode, in which LLM dynamically determines the next speaking agent according to the task context. To support this mechanism, the system organizes all agents into a fixed data structure during initialization, allowing the model to efficiently select from the candidate set.

In existing MAS development frameworks, agents are usually defined in a highly encapsulated manner, including specifications such as the agent's name, prompt, and intended function. These descriptions are typically expressed in natural language. The selected agent then executes actions~\ding{175} based on its role definition and long-term memory~\ding{173}. Such actions generally involve tool invocation, component execution, or communication with other agents. The interaction process continues until a termination mode is triggered, which is based on different types.  
A classic type is based on keyword ending, once  a keyword appears, the responsible agent autonomously ends the process and outputs the final result to the user~\ding{177}.

\section{FLARE: Fuzzing LLM-Based Multi-Agents for Revealing Errors}


\subsection{Overview}

\begin{figure*}
    \centering
    \includegraphics[width=\textwidth]{image/overview15.png}
    \caption{An overview of \tool.}
    \label{3.1}
    \vspace{-10pt}
\end{figure*}

In this paper, we propose \tool, a coverage-guided testing framework for MAS.
As illustrated in Figure~\ref{3.1}, \tool contains three phases: (i) software analysis, (ii) fuzzing testing, and (iii) failures identification.

In the software analysis phase, \tool establishes the ground truth and testing boundaries. By ingesting the System Under Test (SUT) source code and domain-specific knowledge, the \textit{Specification Agent} and \textit{Space Agent} synthesize two critical artifacts: (1) Specifications, which define the expected behavioral specifications; and (2) the behavioral space, which maps the theoretical boundaries of valid agent interactions to serve as the denominator for coverage calculation.

In the fuzzing loop phase, \tool drives the systematic exploration of the MAS state space. We initialize a seed corpus from the framework's predefined tasks and iteratively apply mutations to agent configurations and execution topologies. \tool executes these mutated seeds and monitors runtime traces to compute coverage metrics. Adopting a coverage-guided feedback mechanism, seeds that unlock new states (\ie, increase coverage) are prioritized, dynamically optimizing the selection and mutation strategies for subsequent rounds.

In the failure identification phase, \tool builds oracles based on the specifications to identify semantic deviations by cross-referencing execution logs against the generated specifications. To ensure report reliability, we employ a dual-agent verification mechanism. The \textit{Failure Agent} first scans for potential specification violations. Subsequently, a \emph{Judge Agent} performs a secondary review of these findings to verify their validity, effectively mitigating false positives caused by LLM hallucinations before generating the final failure report.

\subsection{Software Analysis}\label{bd}
To capture the semantics of task completion in a MAS, \tool classifies system behaviors into two distinct categories: intra-agent behaviors and inter-agent behaviors. The formal definitions are as follows:

\emph{Intra-agent behaviors}. We define intra-agent behaviors as the internal decision-making and action processes of a single agent during task execution. In MAS, each agent is treated as an independent entity that makes decisions based on its prompt-specified objectives and its perceived environment state. An intra-agent behavior thus consists of the sequence of outputs and external tool invocations produced by the agent to accomplish its assigned task, independent of interactions with other agents.

\emph{Inter-agent behaviors}. We define inter-agent behaviors as the dynamic processes of interaction among multiple agents at the system level. An inter-agent behavior is represented as a time-evolving sequence of interaction events that is organized by dependency relationships among agents (\eg, ordering constraints across their actions), and the process evolves from the initiation of interaction to a well-defined terminal pattern at the system level.

\subsubsection{SUT specification Generation}\label{3.21} 

\begin{figure}
    \centering
    \includegraphics[width=0.8\linewidth]{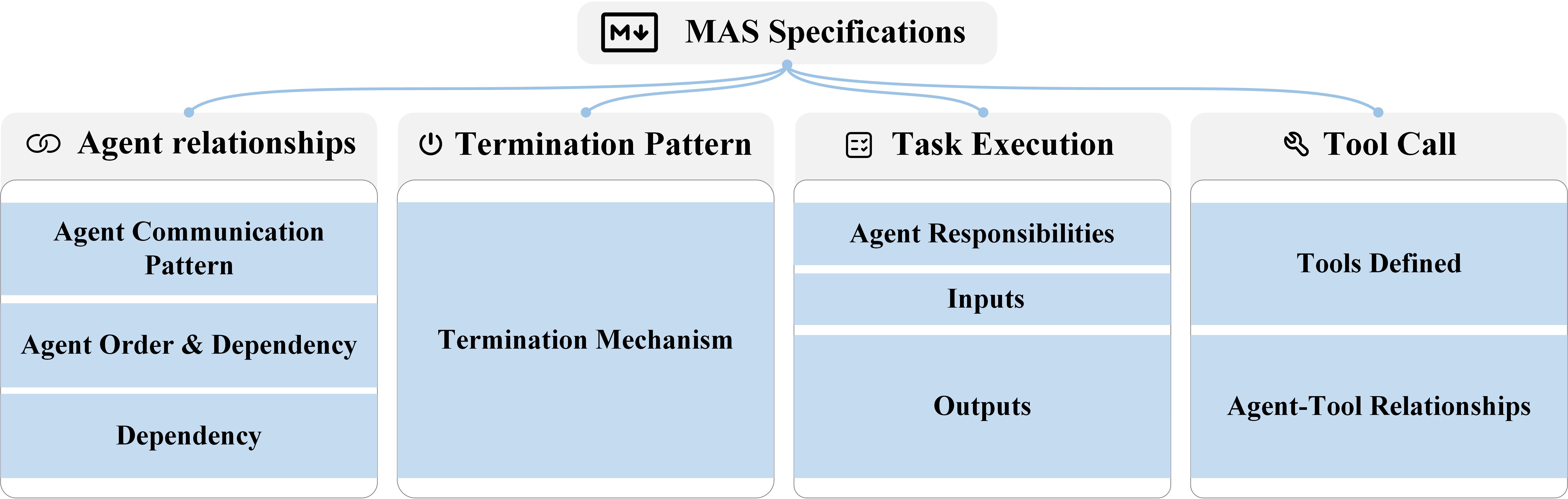}
    \caption{The structure of MAS specification.}
    \label{fig:content}
    \vspace{-10pt}
\end{figure}

The testing specification is designed to formalize the expected behavioral specifications of the SUT and serves as the criterion for validating the correctness of MAS execution. Since the correctness of a MAS depends both on how each agent executes its assigned tasks and on how multiple agents coordinate at the system level, we leverage the behavioral categorization introduced in Section~\ref{bd} to structure the specifications. As depicted in Figure~\ref{fig:content}, the specification schema concretizes these behaviors into four parts:  agent relationships and termination patterns of inter-agent, alongside task execution and tool calls of intra-agent level.

\tool designs a \emph{Specification Agent} to extract specifications from the definitions in the agents' initialization code. It focuses on system prompts, descriptions,  configuration attributes, and the system's communication patterns. To effectively reason about these structures, it utilizes the Chain-of-Thought (CoT) technique, underpinned by comprehensive domain knowledge regarding agent construction and communication mode.

We carefully design the prompts to enable the Specification Agent to internalize this knowledge. Unlike vanilla LLMs, which may hallucinate or misinterpret framework-specific logic, \tool explicitly integrates the framework's operational rules as prior knowledge. This knowledge is rigorously sourced from the development framework's specifications, transforming abstract parameters into actionable reasoning constraints. The prompt establishes a clear domain knowledge mapping, guiding the model to derive specifications that are logically consistent with both the code definitions and the framework's inherent mechanisms.

Specifically, the prompt consists of five key components (details in supplementary material).

    \begin{itemize}
        \item \textbf{Input:} SUT Code, domain knowledge of the MAS framework, and software input specifications. 
    
        \item \textbf{Reasoning Chain:} A three step reasoning chain. (1) Framework-level semantic abstraction: identifying the parameter structures and communication semantics from the underlying MAS framework domain knowledge.
        (2) SUT-specific logic extraction: harvesting granular agent configurations (\eg, prompts and tool schemas) and communication modes (\eg, transition rules and termination patterns) from the application source code; 
        (3) Cross-domain specification synthesis: integrating framework semantics with application-specific logic to generate final behavioral specifications.
    
        \item \textbf{Generation Rules:} The generation guidelines of four categories of system behavior specifications: (1) Agent relationships, which involve the identification of communication pattern, the direct mapping of workflow rules to corresponding sequences, and the recognition of agent dependencies in free-form dialogues (with specific focus on dependencies caused by agent task requirements, system task ordering, and termination conditions); (2) Termination pattern, referring to the specification of how system interaction concludes; (3) Agent tasks, entailing the definition of individual task inputs, outputs, and responsibilities; and (4) Tool invocation, covering the description of tool parameters and agent-tool interactions.
        
        \item \textbf{Example:} A brief introduction to the ShortsMaker~\cite{autogen_workflow_gswithjeff} system in Figure~\ref{fig:1.1} and its corresponding specifications, serving as a one-shot example.
        \item \textbf{Output Format:} The output format is a standardized JSON schema, specifying agent-relationship constraints, system termination conditions, task execution, and tool-call expectations.
    \end{itemize}

\begin{figure}
    \centering
    \includegraphics[width=0.9\linewidth]{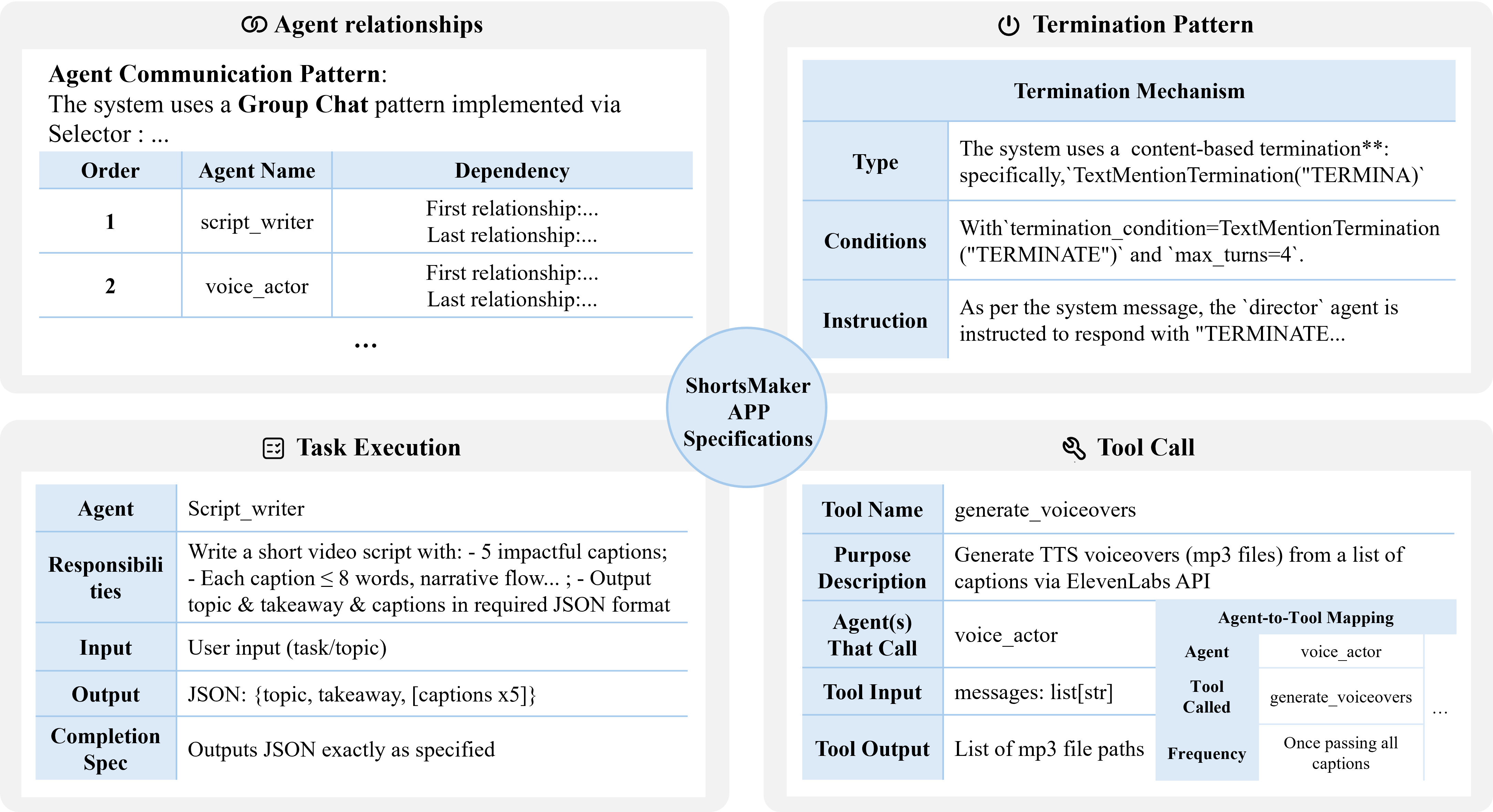}
    \caption{ShortsMaker~\cite{autogen_workflow_gswithjeff} specifications extracted by \tool.}
    \label{fig:oracle}
    \vspace{-5pt}
\end{figure}

\medskip
\emph{Example.}
Consider the SUT shown in Figure~\ref{fig:1.1}. The \tool-extracted specification is illustrated in Figure~\ref{fig:oracle}, including the expected inputs, outputs, and task summaries of different tasks of \textit{director} to help subsequent failures identification. For instance, the \textit{director} is equipped with two task expectation specifications: video generation and terminating the workflow. That is, after the image and audio generation is completed by the \textit{graphic designer agent} and \textit{voice actor agent}, an external API should be invoked to perform the video generation task. The workflow termination task will be executed only when the video has been successfully generated.

\subsubsection{Behavioral Space Generation}\label{sec:3.1}

To construct the MAS behavioral space, \tool builds two complementary abstractions corresponding to the behavioral definitions in Section~\ref{bd}: the intra-agent behavioral space and the execution path space.

\textbf{Intra-agent behavioral space construction.} \tool constructs the intra-agent behavioral space by analyzing developer-provided prompt templates and tool descriptions. The \emph{Space-Agent} module parses these inputs to identify semantically distinct requirements, segmenting them into atomic tasks based on two granularity principles: treating different choices within the same context as separate tasks, and distinguishing choices across different contexts. Beyond the expected behaviors derived from these rules, the space explicitly incorporates three types of MAS anomalies to rigorously evaluate system boundaries. These include empty utterances, where an agent produces no valid output for three consecutive turns; Unproductive loops, in which interactions persist fruitlessly after termination conditions are met; and objective deviation, characterized by an agent's outputs semantically drifting beyond its assigned scope.

\textbf{Execution path space construction.}
\tool constructs the execution path space by enumerating all valid interaction sequences among agents under the given communication patterns. \emph{Space-Agent} first extracts communication patterns from the MAS framework domain knowledge and the SUT code, identifying sender---receiver relationships, message types, and triggering conditions. To address the diverse communication modes introduced in Section~\ref{sec2}, \emph{Space-Agent} incorporates execution path generation schemes within its prompt, enabling differentiated processing tailored to specific pattern characteristics. For systems with workflow-based communication patterns, \emph{Space-Agent} directly instantiates execution paths by following the predefined rules. For systems with free-form communication patterns, \emph{Space-Agent} additionally analyzes both system-level objectives and agent-level task dependencies. Then regards each agent as a node, all paths with orderly node generation space are built according to the system task requirements. 

\begin{figure}
    \centering
    \includegraphics[width=0.7\linewidth]{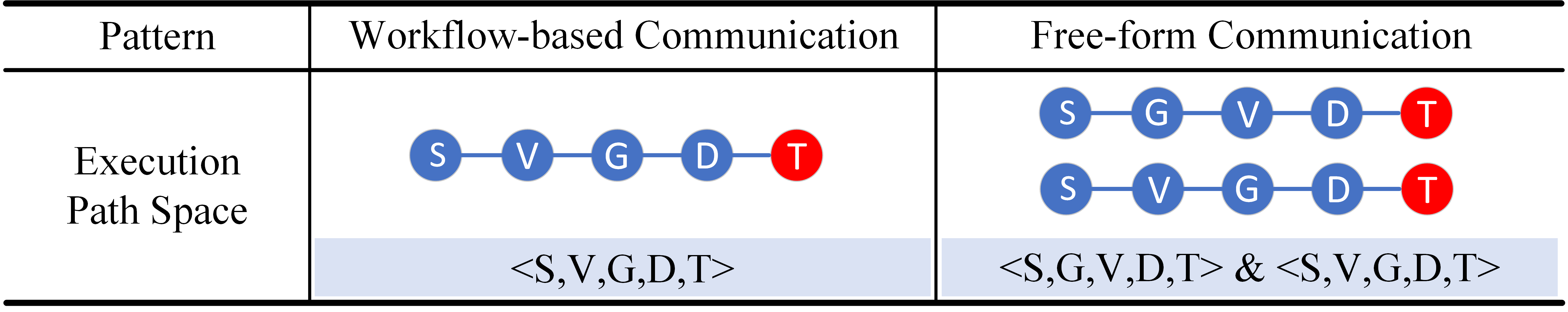}
    \caption{The ShortsMaker~\cite{autogen_workflow_gswithjeff} execution path space of different patterns(S: Script writer; V: Voice actor; G: Graphic designer; D: Director).}
    \label{fig:space}
    \vspace{-5pt}
\end{figure}

\medskip
\emph{Example.}
As illustrated in Figure~\ref{fig:space}, ShortsMaker exhibits distinct execution path spaces based on the communication pattern. The workflow-based pattern imposes a sequential order, resulting in a single valid path. In contrast, the free-form pattern allows the voice actor and graphic designer to process the script writer's output concurrently; the director synthesizes the video only after both complete their tasks. Consequently, the voice actor and graphic designer can swap execution orders, forming two valid paths.

The construction of the behavioral space for \emph{Space-Agent} is based on the comprehensive extraction of SUT source code and framework domain knowledge regarding communication patterns and agent definitions. This process remains consistent with \emph{Specification-agent} in both input data and reasoning logic. Consequently, as shown in the supplementary material, their prompt designs share identical ``Input'' and ``Reasoning Chain'' modules, while the core divergence resides in the ``Instruction'' component, encompassing customized generation rules, few-shot examples, and specific output format constraints.

The prompt of \emph{Space-Agent} ``Instruction'' module designs as follows: 

\begin{itemize} 
    \item \textbf{Generation Rules:} (1) Intra-agent behavioral space Generation, involving the accurate identification of agents within the system under test (based on initiating chat, giving tasks, and human input modes) and the designation of the agent ``name'' attribute as a unique identifier; and the systematic segmentation of agent tasks, entailing the numerical division of tasks (\eg, 1, 2...) and the integration of matching tool invocations into the task content. (2) Execution path space generation, comprising the construction of valid interaction paths by modeling identified agents as nodes connected according to communication modes; the extraction of prescribed speaking sequences governed by fixed workflow rules; the enforcement of maximum turn constraints (truncating paths where node length exceeds limits); and the derivation of paths in free-form modes, covering the extraction of direct invocation relationships or the generation of all potential permutations based on task logic when explicit dependencies are absent.
    \item \textbf{Example:} Similarly, the presentation of the application overview, the workflow for the construction of the corresponding generation space, and the resulting space, with the Shortsmaker\cite{Gswithjeff2024Workflow} system serving as a running example for intra-agent behavioral space generation and workflow-based execution path space generation, and Italian Estate Analyst\cite{Josephrp2024RealEstate} illustrating execution path space generation under free-form modes.
    \item \textbf{Output Format Requirements:} The adoption of a unified JSON representation, comprising a single-agent behavioral space of multiple behaviors (each specified by the agent name, its class as defined in the framework code, and a textual behavior description) and an execution path space of multiple complete paths (each represented as an ordered list of agent names). 
\end{itemize}

\subsection{Fuzz Testing}

\subsubsection{Fuzzing Loop Overview}
\tool designs a MAS-specific testing approach based on the classic fuzzing loop \cite{liang2018fuzzing}. Targeting MAS characteristics, \tool develops a customized seed pool, a weighted random seed selection strategy, and a mutator guided by intra-agent and inter-agent coverage metrics. The fuzzing loop repeatedly selects a seed from the corpus, mutates it to obtain a new test input, executes the SUT, and then updates coverage and the seed corpus.

\subsubsection{Seeds Generation and Selection}

At the onset of the testing phase, \tool initializes the seed pool as follows:
\begin{equation}
    \mathcal{R} = \left\{ (I_i, C_i, S_i) \right\}_{i=1}^{N}, \quad \mathcal{W} = \{ w_i \}_{i=1}^{N}
\end{equation}

To commence the initial testing round, \tool constructs a baseline set $R$ of seeds. We prompt an auxiliary LLM with the SUT documentation to generate $N-1$ task descriptions, and we additionally include a null input that leaves the initial state unchanged. For these initial seeds, the model configuration $C_i$ is standardized according to the SUT code, and the sequence $S_{a_1,...,n}$ is initialized to match the default agent arrangement defined within the SUT(n represents the number of agents in the SUT).

\tool uses a classic weighted random selection strategy\cite{jiang2009adaptive} to select these seeds. Initially, identical weights $W$ are assigned to all seeds ($w_i = w_{init}$). Subsequently, these selection weights are dynamically adjusted in response to coverage variations until they reach predefined upper or lower thresholds, thereby optimizing the exploration of the state space.
\subsubsection{Seeds Mutation}

\tool performs bursts on variable parameters within a seed after selection to increase the diversity of test cases. We define the System Input $I$ as an invariant during the mutation phase to preserve the validity of the SUT specification. Mutating the semantic content of $I$ (\eg, the user requirement or task description) would alter the ground truth of the task, rendering the predefined validation specifications inapplicable. For each selected seed, \tool generates 1 mutated variant. For each variant, \tool independently applies configuration mutation and sequence mutation, so that a test case may change the model configuration $C$, the initial sequence $S$, or both. Notably, if the MAS enforces a fixed communication pattern defined by the developers, sequence mutation is bypassed to preserve the intended interaction logic.

\textbf{Strategy 1: Configuration Mutation.}
For the model configuration $C$, \tool employs a hierarchical mutation strategy that first randomly designates a single agent within the SUT as the target. Subsequently, the system applies one of four distinct mutation operators: Identity (no change), Temperature Scaling, Model Family Switching, or Joint Mutation, the latter involving simultaneous alterations to both temperature and model family. To optimize efficiency, these operators are selected using an adaptive weighted sampling approach, where specific weights are iteratively tuned based on coverage gains and clamped by predefined thresholds.

\textbf{Strategy 2: Sequence Mutation.}
Regarding the initial sequence $S_{a_1,...,n}$, \tool first identifies the communication pattern of the SUT. The sequence mutation is exclusively activated for SUTs exhibiting the free-form interaction pattern. In this mode, \tool adopts a random selection mutation strategy, randomly selecting a single sequence from the full agent permutation for mutation. Conversely, for workflow-based patterns, the sequence remains immutable. This constraint is imposed because the initial sequence in workflows typically encodes a rigid execution logic; forcing a reordering would disrupt the SUT's structural integrity, resulting in semantically invalid test cases that do not reflect the system's intended behavior.

\subsubsection{Execution and Coverage Feedback} \label{cf}

During the execution of mutated seeds on the SUT, \tool collects both the full execution log and a semantic event. The full log, which preserves all raw messages, reasoning chains, and tool invocations, is utilized for coverage measurement. Its semantic completeness ensures that the LLM can accurately map subtle execution contexts to behavior definitions without information loss. Conversely, the semantic event sequence serves as the primary basis for failure identification. It comprises a chronological trace of agent executions, structured tool records (detailing caller identity, tool nomenclature, and output data), and condensed utterances, specifically the initial, median, and terminal sentences of a dialogue. This structured abstraction effectively filters out contextual noise, enabling the Specification to focus on the identification of functional failures within a reduced data volume.

To quantify the exploration of the MAS, \tool implements two specific coverage criteria (the construction of these behavioral spaces is detailed in Section~\ref{sec:3.1}).
The first metric, \textbf{Intra-agent behavior coverage}, assesses the extent to which agent actions explore the total behavioral space defined in Section~\ref{sec:3.1}. Formally, based on this space, \tool assigns unique identifiers to each agent and behavior, constructing a sparse matrix where row indices correspond to agent ids and column indices map to behavior ids. To quantify coverage, the system implements a semantic-to-identifier mapping mechanism. During execution, \tool utilizes an LLM to semantically align the raw execution context with candidate behaviors. Through this analysis, the model resolves the semantically matched behavior into its corresponding behavior id, which is subsequently used to index and update the respective cell in the sparse matrix. 

The second criterion, \textbf{inter-agent behavior coverage}, quantifies how well the execution traces cover all legal interaction paths. \tool initially performs a direct check of the agent invocation sequence against the legal paths defined in Section~\ref{sec:3.1}. To handle complex interactions, \tool employs relaxation rules-including merging consecutive utterances, inlining nested tool calls, and reordering independent steps-prior to re-matching. Similar to the intra-agent analysis, the matching of execution sequences and utterances to legal paths is delegated to an LLM, and the final coverage is computed as the ratio of covered paths to the total number of legal paths.

These two coverage criteria are jointly utilized to guide the fuzzer's seed selection and mutation processes through a unified probability adjustment mechanism. Rather than evaluating the metrics independently, \tool assesses whether a test case increases the cumulative coverage across either criterion. If the cumulative coverage increases, the selection probability of the parent seed and its associated mutation strategy is incremented by a fixed step size; conversely, if the coverage stagnates or declines, the probability is decremented.

\medskip

\begin{figure}
    \centering
    \includegraphics[width=0.9\linewidth]{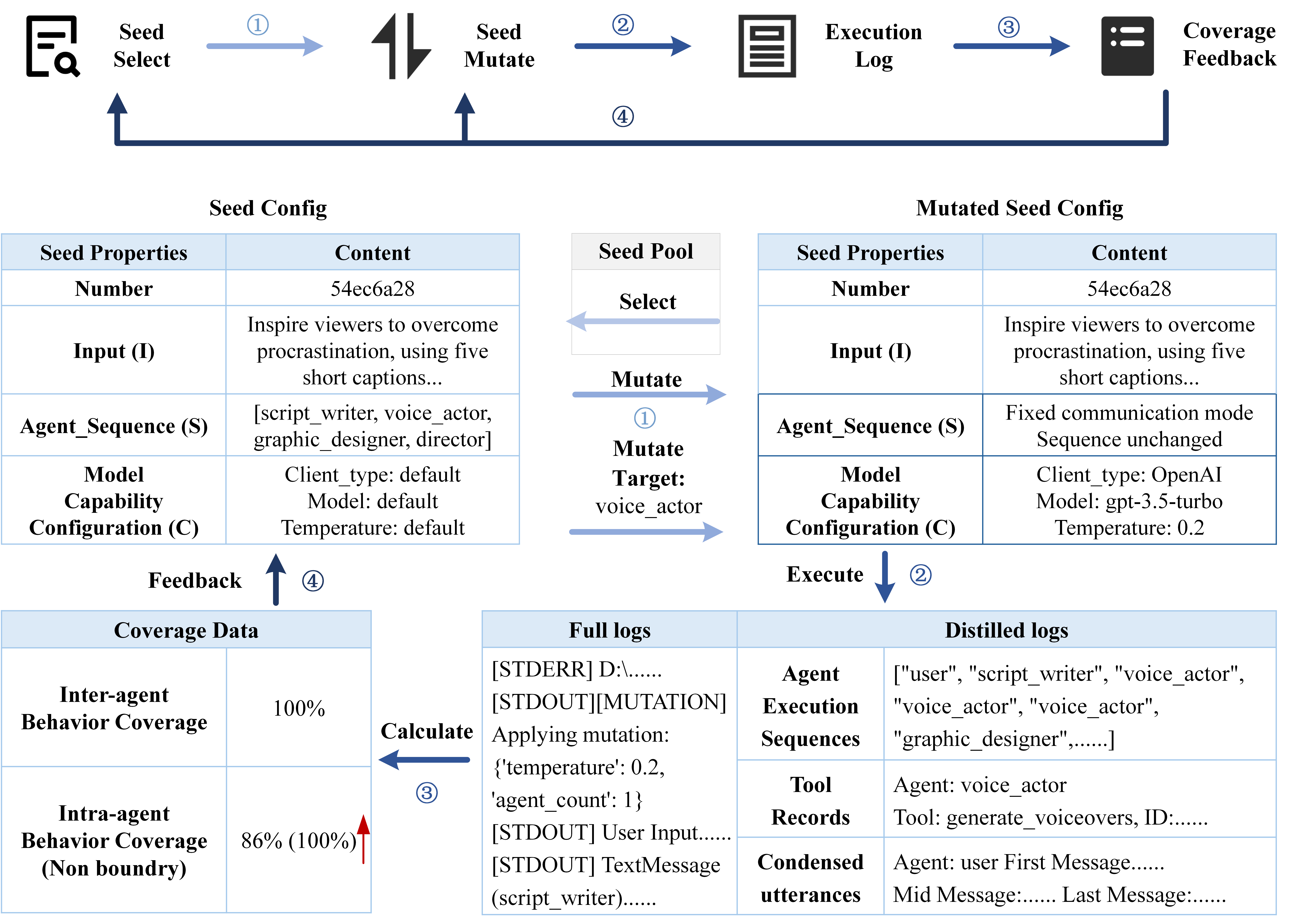}
    \caption{The Test execution of ShortsMaker by \tool.}
    \label{fig:test}
    \vspace{-5pt}
\end{figure}

\emph{Example.} Figure~\ref{fig:test} illustrates the execution process of a test case on ShortsMaker~\cite{autogen_workflow_gswithjeff}. \tool first selects a seed from the pool. As the SUT operates with the workflow-based pattern, \tool applies mutation strategy 2 (capability mutation). Following execution, \tool extracts two types of structured logs and computes the current coverage using the full log. For inter-agent behavior, which contains only a single legal path in this fixed mode, \tool confirms coverage after merging execution traces and applying relaxation rules. Regarding intra-agent behavior, the analysis reveals that all normal behaviors are covered, while two boundary behaviors remain uncovered. Finally, the improvement in intra-agent behavior coverage is fed back to increment the selection weights of both the seed and the mutation strategy.

\subsection{Failures Identification}

In this part, \tool builds oracles based on the specifications defined in Section~\ref{3.21}. Through these oracles, \tool detected four types of failure symptoms: agent task execution failures, tool invocation failures, agent dependency failures, and system termination failures.

\textbf{Agent task execution failures.} \tool reports a task execution failure if (1) the agent produces results inconsistent with the defined task outputs (\eg, structurally invalid or semantically incorrect); (2) the agent performs actions exceeding its assigned responsibilities; or (3) the agent explicitly refuses to fulfill its responsibilities.

\textbf{Tool invocation failures.} \tool reports a tool misuse failure if (1) the agent invokes a tool that violates the identified agent-tool interactions; or (2) the invocation is inconsistent with the described tool parameters.

\textbf{Agent relationships failures.} \tool reports a relationship failure when either (1) a deviation from the corresponding sequences defined by fixed rules, or (2) a violation of agent dependencies in free-form dialogues (specifically regarding system task ordering or termination conditions).

\textbf{System termination failures.} \tool reports a termination failure when the system fails to adhere to the specification of how system interaction concludes. \tool focuses on two distinct termination anomalies: (1) the failure to conclude operations, where agents persist in exchanging messages or performing mutual invocations even after the termination conditions are already satisfied; and (2) premature termination, where the system erroneously halts the interaction before the required criteria are met.

\medskip

\tool identifies defects through the workflow between the failure Agent and the Judge Agent. These two agents share the log, specifications, and system context. One agent directly detects the log against the specification for the four types of error symptoms, while the other agent verifies the correctness of the first agent's findings. Through multi-turn interaction, the two agents reach a consensus---or exit after a maximum number of rounds---and produce a defect analysis report. Specifically, we design a dedicated prompt template in which both agents share the same input and reasoning chain, but use customized instructions for their respective roles.

  \begin{itemize}
            \item \textbf{Input:} It includes the specification produced by the specification agent as the reference specification of expected behavior, and log files collected specifically for error diagnosis. There are four types of logs, including \emph{agent speaking order}, containing the actual turn sequence of agent utterances during execution; \emph{tool-call log}, containing detailed records of each tool invocation; \emph{per-agent utterance excerpts}, containing excerpts from the beginning, middle, and end of each agent's messages; \emph{dead-loop test result}, containing a Boolean flag indicating whether the log exhibits timeout or cyclic dialogue behavior.
            
            \item \textbf{Judge Analysis Chain:} A three-step process: (1) \emph{Specification-log distinction:} identifying discrepancies between specifications and execution logs regarding agent behaviors; (2) \emph{Logic verification:} validating consistency between log information and Specification constraints; and (3) \emph{Structured output generation:} formatting the final judgments according to the required template. 
            
            \item \textbf{Instruction of Failures Agent:}
            The \emph{requirement} for the agent to utilize four error categories---agent task execution errors, tool invocation errors, agent relationships errors, and system termination errors---as detection rules for the identification of faults in execution logs. This includes the presentation of an illustrative failure-identification example on the Business Challenge Resolver application\cite{agentcy} for detecting an agent relationships error, comprising the provision of the faulty log segment, the expected agent relationships specification, and the final violation judgment. Furthermore, the mandatory reporting of violations in a JSON object with two fields: the violated-specification type and a description specifying the offending log segment and the reason for the violation (or the return of an empty value if no violation is found).

            \item \textbf{Instruction of Judge Agent:}
            A verification directive designed for the secondary review of reported violations. It comprises logic for localizing failure-inducing log segments to detect omissions and misjudgments, and defines the criteria for synthesizing violation reports, raw logs, and the Business Challenge Resolver to determine a final \textbf{CORRECT} or \textbf{INCORRECT} verdict.
        
    \end{itemize}
\section{Evaluation}
 We conducted our evaluation to answer the following questions:
\begin{itemize}
    \item \textbf{RQ1 (Specification):} Does \tool correctly generate MAS specifications?
    \item \textbf{RQ2 (Coverage):} Does \tool help improve various coverage metrics in testing?
    \item \textbf{RQ3 (Failures):} How effective is \tool in identifying MAS unique failures?
\end{itemize}



\subsection{Methodology}

\subsubsection{Applications}
We use AutoGen~\cite{microsoft_autogen_github}, the most popular framework (52.4K stars by Oct 2025) specialized in MAS development, as our primary research platform. To ensure the representativeness and quality of our benchmark, we surveyed application repositories on GitHub using the keywords ``autogen-extension and autogen application'', which is officially recommended by the AutoGen documentation for identifying community contributions. From the search results, we selected the top-15 projects ranked by star count, supplemented by one official example application. These applications span diverse task scenarios, including automated code generation, video production, data analysis, and deep research (details in Table~\ref{tabe1:part1}). Notably, selected applications incorporate distinct technical components, such as external tool invocation, knowledge retrieval, and nested multi-agent group chats. 

The size of these applications ranges from 80 to 550 Lines of Code. As of the time of this study, the projects have a median age of 13 months, with 15 of the 16 applications having received one or more stars on GitHub. Detailed information for each application, including links to their respective code repositories, is provided in Table~\ref{tabe1:part1}.

\subsubsection{Baselines}
Since there is no prior work specifically targeting logic failures in MAS, we compared \tool against 3 baseline techniques derived from general LLM testing and traditional software fuzzing. The first one targets LLM safety, while the latter two focus on traditional code coverage. As we will see, these baselines generally lack the semantic understanding of agent interactions required for MAS, often failing to trigger deep logic failures. They are adapted here primarily for comparison purposes to demonstrate the necessity of MAS-specific guidance.

\begin{itemize}
    \item \textbf{LLM-Fuzzer}~\cite{299691}: It utilizes a fuzzing loop-incorporating seed selection and mutation to provoke LLMs into exhibiting unintended behaviors. While effective for detecting harmful or biased text in single models, this baseline focuses on safety alignment and does not account for the complex, multi-turn interaction logic or task-completion goals inherent in MAS. Although originally designed for prompt jailbreaking, in this paper, we adapt this method to mutate user inputs, thereby generating a diverse range of initial queries.
    
    \item \textbf{PythonFuzz}~\cite{pythonfuzz_github}:
    It is a coverage-guided fuzzing framework that ports the logic of American Fuzzy Lop (AFL)~\cite{afl_github} to Python. It utilizes instrumentation to track execution paths and employs random mutation strategies (\eg, bit flips, byte shuffling) to generate inputs. This baseline treats inputs strictly as byte sequences rather than semantic text, aiming primarily to maximize code coverage and detect uncaught exceptions or crashes in standard Python functions rather than logical deviations in agents.
    
    \item \textbf{Frelatage}~\cite{frelatage_github}:
    It is a coverage-based fuzzer specifically optimized for fuzzing Python applications with structured data. Unlike purely random fuzzers, Frelatage leverages dictionary-based generation and type hints to produce syntactically plausible inputs. It focuses on exploring edge cases in the control flow graph by optimizing the diversity of input arguments.
\end{itemize}

\subsubsection{Metrics}
We employ four test coverage metrics. 

\begin{itemize}
    \item Statement coverage (SC): the portion of executed statements in all executable ones.
    \item Branch coverage (BC): the portion of executed branch in all branches.
    \item Inter-agent behavior coverage (RAC): the portion of executed traces in all execution path space. Details in Section~\ref{cf}.
    \item Intra-agent behavior coverage (AAC) : the portion of the agent boundary and expected behaviors in all intra-agent behavior space. Details in Section~\ref{cf}.
\end{itemize}





\subsection{Answer to RQ1: Specification}

To evaluate the accuracy of the specifications generated by \tool, three of the authors manually examine all the specifications. To mitigate potential bias, we employed a double-blind protocol: two authors independently reviewed the generated oracles without knowledge of their source or the underlying generation logic. Their task was to assess whether each oracle adhered to the functional logic and compliance rules defined in the prompt (Section~\ref{3.21}). Their results are cross-validated, and a third author joins when they encounter consensus problems.

To rigorously quantify the inter-rater reliability, we utilized the \emph{Linear Weighted Cohen's Kappa}\cite{mchugh2012interrater} coefficient. Unlike standard Kappa, this metric accounts for the degree of disagreement by assigning penalties linearly proportional to the difference between ratings (\ie, a discrepancy of 1 unit incurs a penalty of 1). The calculated coefficient was $0.8512$, indicating a strong level of agreement between the raters. In cases of remaining disagreements, participant $x$ provided a final adjudication, maintaining anonymity regarding the prior judgments of $y$ and $z$.

Across all 64 specifications components(each application has 1 specification, 1 specification has 4 components) generated for 16 applications, \tool achieved high alignment with the ground truth, with only three discrepancies. These instances were limited to edge cases involving ambiguous attribute semantics in AutoGen's free-discussion mode. These results confirm \tool's ability to accurately capture system specifications in diverse interaction scenarios.

\subsection{Answer to RQ2: Coverage} 

\begin{table}
    \centering
    \scriptsize 
    \renewcommand{\arraystretch}{1.3} 
    \setlength{\tabcolsep}{2pt} 
    \newcolumntype{C}{>{\centering\arraybackslash}X}
    
    \caption{Overall results of \tool.}
    \begin{tabularx}{\linewidth}{|c|l|CCCC|CCCC|CCCC|} 
        \hline
         \multicolumn{2}{|c|}{\multirow{2}{*}{\textbf{Task \& Application}}} & \multicolumn{4}{c|}{Statement Coverage (SC)} & \multicolumn{4}{c|}{Branch Coverage (BC)} & \multicolumn{4}{c|}{Inter-agent Coverage(RAC)} \\ \cline{3-14}
        
         \multicolumn{2}{|c|}{} & \makecell{\tool} & \makecell{LLM\\Fuzz} & \makecell{Fre\\latage} & \makecell{Py\\Fuzz} & \makecell{\tool} & \makecell{LLM\\Fuzz} & \makecell{Fre\\latage} & \makecell{Py\\Fuzz} & \makecell{\tool} & \makecell{LLM\\Fuzz} & \makecell{Fre\\latage} & \makecell{Py\\Fuzz} \\ \hline
        
        \multirow{4}{*}{\shortstack{\textbf{Report}\\\textbf{Generation}}} 
          & Business Resolver\cite{SCTYInc2024Agentcy} & 83.7 & 61.1 & / & 83.7 & 25 & 20 & / & 25 & 100 & 100 & / & 100 \\ 
          & Financial Reporter\cite{Pratyushkr2024Financial} & 100 & 100 & 94.9 & 100 & 100 & 100 & 50 & 100 & 100 & 100 & 100 & 100 \\ 
          & Recruitment Helper\cite{Castaldo2024Recruitment} & 86.1 & 79.2 & 85.9 & 86.1 & 87.5 & 69.2 & 87.5 & 87.5 & 100 & 100 & 100 & 0 \\
          & Grant Writer\cite{LewisExternal2024GrantWriter} & 75.5 & 36.8 & 37.6 & 71.3 & 36.4 & 22.2 & 17.9 & 22.7 & 100 & 100 & 100 & 100 \\ \hline
        
        \multirow{3}{*}{\shortstack{\textbf{Data}\\\textbf{Analytics}}}
          & Company Researcher\cite{Microsoft2024AutoGen} & 95.1 & 86.6 & 29.7 & 78.2 & 79.2 & 69.2 & 0* & 54.2 & 50 & 50 & 50 & 50 \\ 
          & Italian Estate Analyst\cite{Josephrp2024RealEstate} & 81.5 & 70.4 & 55.5 & 36.4 & 42.6 & 42.6 & 2 & 25.6 & 100 & 100 & 100 & 0 \\ 
          & Tool Caller\cite{Gordu2024Multiagent} & 95.7 & 86.2 & 60.9 & 50 & 83.3 & 66.7 & 83.3 & 83.3 & 100 & 100 & 100 & 100 \\ \hline

        \multirow{2}{*}{\shortstack{\textbf{Role}\\\textbf{Playing}}}
          & Joke Maker\cite{Kingglory2024DesignPatterns} & 100 & 100 & 100 & 100 & 100 & 100 & 100 & 100 & 100 & 100 & 100 & 100 \\ 
          & Customer Onboarding\cite{Kingglory2024DesignPatterns} & 95.5 & 93.9 & 95.8 & 95.5 & 75 & 75 & 75 & 75 & 100 & 100 & 100 & 100 \\ \hline

        \multirow{2}{*}{\shortstack{\textbf{Code}\\\textbf{Generation}}}
          & Financial Analyst\cite{Kingglory2024DesignPatterns} & 100 & 100 & 100 & 100 & 100 & 100 & 100 & 100 & 100 & 100 & 100 & 100 \\
          & Game Creator\cite{Gordu2024Multiagent} & 100 & 100 & 100 & 100 & 100 & 100 & 100 & 100 & 100 & 100 & 100 & 100 \\ \hline

        \textbf{Video Prod.} & Shorts Maker\cite{Gswithjeff2024Workflow} & 93.7 & 92.7 & 27.8 & 64.3 & 77.8 & 81.0 & 5.3 & 52.8 & 100 & 100 & 100 & 100 \\ 
        \textbf{Feedback Map} & Blog/post Writer\cite{Kingglory2024DesignPatterns} & 100 & 100 & 100 & 100 & 100 & 100 & 100 & 100 & 100 & 100 & 63.6 & 100 \\ 
        \textbf{Board Sim.} & Chess Simulation\cite{Kingglory2024DesignPatterns} & 100 & 100 & 100 & 100 & 100 & 100 & 100 & 100 & 100 & 0 & 0 & 0 \\ 
        \textbf{Deep Research} & Workflow Designer\cite{Gordu2024Multiagent} & 100 & 100 & 100 & 100 & 100 & 100 & 100 & 100 & 100 & 100 & 100 & 100 \\ 
        \textbf{Visual Tasks} & Image Explainer\cite{Gordu2024Multiagent} & 100 & 100 & 37.5 & 44.4 & 100 & 100 & 100 & 100 & 100 & 0 & 100 & 0 \\ \hline 
        
        \multicolumn{2}{|c|}{\textbf{Average}} & \textbf{94.2} & 87.9 & 75.0 & 81.9 & \textbf{81.7} & 77.9 & 68.1 & 76.6 & \textbf{96.9} & 84.4 & 87.4 & 71.9 \\ \hline
    \end{tabularx}

    \vspace{10pt} 


\begin{threeparttable}
    \begin{tabularx}{\linewidth}{|c|l|CCCC|ccccc|ccc|} 
        \hline
        \multicolumn{2}{|c|}{\multirow{3}{*}{\textbf{Task \& Application}}} & 
        \multicolumn{4}{c|}{\multirow{2}{*}{Intra-agent Coverage (AAC)(A/N)}} & 
        \multicolumn{8}{c|}{Exposed Failures} \\ \cline{7-14}
        
        \multicolumn{2}{|c|}{} & 
        \multicolumn{4}{c|}{} & 
        \multicolumn{5}{c|}{\tool} & 
        \makecell{LLM\\Fuzz} & \makecell{Fre\\latage} & \makecell{Py\\Fuzz} \\ \cline{3-14}
        
        \multicolumn{2}{|c|}{} & 
        \makecell{\tool} & \makecell{LLM\\Fuzz} & \makecell{Fre\\latage} & \makecell{Py\\Fuzz} & 
        AR & ST & TE & TI & RC & 
        RC & RC & RC \\ \hline 
        
        
        \multirow{4}{*}{\shortstack{\textbf{Report}\\\textbf{Generation}}}
          & Business Resolver\cite{SCTYInc2024Agentcy} & 96.7/100 & 96.7/100 & / & 97.4/100 & 1 & 1 & 4 & 2 & &  & / &  \\ 
          & Financial Reporter\cite{Pratyushkr2024Financial} & 95/100 & 86.7/100 & 92.3/100 & 94.1/100 & & 2 & 3 & 1 & 1 & 1 & & 1 \\ 
          & Recruitment Helper\cite{Castaldo2024Recruitment} & 88.9/100 & 85.7/100 & 66.7/100 & 75/100 & & & 1 & 1 & &  &  &  \\
          & Grant Writer\cite{LewisExternal2024GrantWriter} & 83.3/100 & 88.9/100 & 88.9/100 & 71.4/80 & 1 & & 2 & & & &  &  \\ \hline
        
        \multirow{3}{*}{\shortstack{\textbf{Data}\\\textbf{Analytics}}}
          & Company Researcher\cite{Microsoft2024AutoGen} & 85.7/100 & 92.3/100 & 80/100 & 81.8/100 & & 1 & 2 & 2 & & 1 &  & \\ 
          & Italian Estate Analyst\cite{Josephrp2024RealEstate} & 93.3/100 & 94.1/100 & 86.7/100 & 21.1/23.5 & & & 2 & 2 & 1 &  &  &  \\ 
          & Tool Caller\cite{Gordu2024Multiagent} & 100/100 & 100/100 & 100/100 & 87.5/100 & & 2 & & 1 & &  &  &  \\ \hline

        \multirow{2}{*}{\shortstack{\textbf{Role}\\\textbf{Playing}}}
          & Joke Maker\cite{Kingglory2024DesignPatterns} & 77.8/100 & 75/100 & 66.7/100 & 90/100 & & & & & &  &  &  \\ 
          & Customer Onboarding\cite{Kingglory2024DesignPatterns} & 89.5/100 & 100/100 & 90/100 & 71.4/100 & & 1 & 2 & & &  & 1 &  \\ \hline

        \multirow{2}{*}{\shortstack{\textbf{Code}\\\textbf{Generation}}}
          & Financial Analyst\cite{Kingglory2024DesignPatterns} & 91.7/100 & 90/100 & 66.7/60 & 90/100 & & 1 & 2 & 1 & 1 & 1 &  &  \\
          & Game Creator\cite{Gordu2024Multiagent} & 100/100 & 90.9/100 & 100/100 & 87.5/100 & & & & & & 1 & 2 &  \\ \hline

        \textbf{Video Prod.} & Shorts Maker\cite{Gswithjeff2024Workflow} & 85.7/100 & 85.7/100 & 66.7/100 & 66.7/100 & & 1 & 1 & 2 & & 1 &  &  \\ 
        \textbf{Feedback Map} & Blog/post Writer\cite{Kingglory2024DesignPatterns} & 93.3/100 & 94.4/100 & 80/100 & 81.8/100 & 1 & & 1 & & &  & &  \\ 
        \textbf{Board Sim.} & Chess Simulation\cite{Kingglory2024DesignPatterns} & 93.3/100 & 92.9/100 & 85.7/100 & 83.3/100 & 1 & 1 & 1 & 2 & 1 &  &  & 1 \\ 
        \textbf{Deep Research} & Workflow Designer\cite{Gordu2024Multiagent} & 100/100 & 93.3/100 & 94.7/100 & 94.7/100 & & 1 & 2 & 2 & 1 & 1 &  &  \\ 
        \textbf{Visual Tasks} & Image Explainer\cite{Gordu2024Multiagent} & 88.9/100 & 75/100 & 66.7/100 & 71.4/100 & & & 2 & & &  & 1 &  \\ \hline
        
        \multicolumn{2}{|c|}{\textbf{Average (Total Failures)}} & \textbf{91.4}/\textbf{100} & 90.1/100 & 82.1/97.3 & 79.1/94.0 & 4 & 11 & 25 & 16 & 5 & 6 & 4 & 2 \\ \hline
    \end{tabularx} 
    \begin{tablenotes}
            \footnotesize 
            \item[*] The zero coverage results from concurrency deadlocks triggering Frelatage's timeout-based forced termination, which precludes the saving of execution logs.
            \item[*] A/N: Intra-agent coverage calculated over the entire behavior space (A) vs. excluding boundary behaviors (N).
            \item[*] AR: Agent Relationship; ST: System Termination; TE: Task Execution; TI: Tool Invocation; RC: Runtime Crash.
    \end{tablenotes}
\end{threeparttable}
\label{tabe1:part1}
\end{table}

Table~\ref{tabe1:part1} summarizes the overall results. Due to irreconcilable environmental conflicts, we omit the results of Frelatage on applications.


Across diverse applications, $\text{\tool}$ consistently outperforms all baseline methods in terms of coverage. Specifically, regarding traditional metrics, $\text{\tool}$ achieves the highest SC ($94.2\%$) and BC ($81.7\%$). It is worth noting that the uncovered branches are primarily attributed to either redundant error-handling blocks that remain unreachable under valid execution logic, or specialized branches designed for edge-case anomaly handling (\eg, specific symbols or numerical outliers), which are rarely triggered in standard functional scenarios. In terms of RAC, \tool achieves substantial improvements by utilizing task inputs and initialization sequences to traverse paths elusive to baselines. Notably, in the Chess Simulation\cite{Kingglory2024DesignPatterns}, \tool uniquely captures the correct interaction dynamics; its precise handling of turn-taking ensures the valid execution sequence required for proper termination. Finally, regarding the AAC metric, we observed no significant difference between \tool and the baselines. We attribute this phenomenon to the nascent stage of current MAS development, where the prevalence of simplified, single-task agent designs allows even standard execution to satisfy basic behavioral coverage.

LLM-Fuzzer significantly outperforms Frelatage and Pythonfuzz across both traditional and MAS-specific coverage metrics, although it falls slightly short of Frelatage in RAC. Traditional Python testing methods(Frelatage and Pythonfuzz) rely on byte-level mutations to generate non-semantic noise, rendering the program susceptible to formatting failures that sever inter-agent interactions. In contrast, LLM-Fuzzer generates diverse user inputs by leveraging high-level fuzzy intents rather than raw byte manipulation.

\subsection{Answer to RQ3: Failures}
\begin{table}
    \centering
    \caption{Root causes of the identified failures.}
    \resizebox{\columnwidth}{!}{%
        \begin{tabular}{llcccc}
            \toprule
            Failure type & Root Cause & \tool & LLM-Fuzzer & Frelatage & PythonFuzz \\
            \midrule
            
            Runtime Crash & / & 5 & 6 & 4 & 2 \\
            \midrule
            
            \multirow{6}{*}{Agent Task Execution failures} 
            & Prompt and Instruction Deviation & 15 & - & - & - \\ 
            & Response Omission & 4 & - & - & - \\ 
            & Malformed Output Format & 2 & - & - & - \\ 
            & Repetitive Execution of Completed Tasks & 2 & - & - & - \\ 
            & Role Misalignment & 1 & - & - & - \\ 
            & Explicit Task Refusal & 1 & - & - & - \\ 
            \midrule
            
            \multirow{3}{*}{Tool Invocation failures} 
            & Tool Invocation Omission & 5 & - & - & - \\
            & Tool Execution Error & 7 & - & - & - \\
            & Tool Specification Mismatch & 4 & - & - & - \\
            \midrule
            
            \multirow{1}{*}{Agent relationships failures} 
            & Violation of relationships-based speaking order & 4 & - & - & - \\
            \midrule
            
            \multirow{2}{*}{System Termination failures} 
            & Termination Condition Violation & 5 & - & - & - \\
            & Max Round Limit Exceeded & 6 & - & - & - \\
            \midrule
            \multirow{1}{*}{\textbf{Sum}}  & /&\textbf{61} & \textbf{6} & \textbf{4} & \textbf{2} \\
            \bottomrule
        \end{tabular}%
        }
        \label{bug}
\end{table}

Table~\ref{bug} illustrates that \tool exposed a total of $61$ failures across 16 apps. Only 5 of them lead to software crashes, and the others are functional failures triggered by complex MAS behavior. However, the baselines are limited to detecting only runtime crashes. LLM-Fuzzer detected one additional crash compared to \tool. This specific failure was caused by a context window overflow triggered by its prompt injection attacks, a scenario distinct from the agent behavior failures targeted by our approach.

\textbf{Agent task execution failures.}
Among the 25 task execution failures exposed by \tool, the majority (15) are attributed to Prompt and Instruction Deviation, where agents fail to adhere to specific constraints specified in the prompts. A representative case occurred in \textit{Financial Reporter}\cite{Pratyushkr2024Financial}, where an agent explicitly instructed to analyze real estate listings in Rome erroneously processed data for Milan. The remaining 10 failures involve procedural or formatting anomalies-including response omission (4), malformed output (2), repetitive execution (2), role misalignment (1), and explicit refusal (1)-reflecting general stability issues where agents break interaction protocols or deviate from their designated operational logic.

\textbf{Tool invocation failures.}
Among the 16 tool invocation failures identified by \tool, the most prevalent cause is tool execution failures (7), which occur when the system lacks robust error-handling mechanisms to recover or continue execution after a runtime tool failure. A representative example involved an Arxiv survey task\cite{Gordu2024Multiagent} where the \emph{Engineer} agent generated a scraping script requiring a non-existent dependency; the resulting execution failures caused the \emph{UserProxy Agent} to crash, leading to complete task failure due to the absence of a fallback strategy. The remaining cases involve tool invocation omission (5) and Tool specification mismatch (4), representing instances where agents neglected necessary tool usage or generated arguments that violated the predefined input/output schemas.

\textbf{Agent relationships failures.}
All 4 instances of agent relationship failures are classified as speaking order violations. This failure mode arises when an agent is invoked out of its predefined dependency sequence, disrupting the logical data flow required for task completion.

\textbf{System termination failures.}
\tool detects failures categorized into termination condition violation (6) and max round limit exceeded (4). The former stems from defective termination logic where the preset stop conditions fail to trigger or conflict with global constraints, preventing a natural conclusion. The latter results from improper configuration of the execution budget, where the maximum round limit is set too low relative to the task complexity, causing premature termination before objectives are met.




\subsection{Impact of underlying LLMs}

\begin{table}[!ht]
    \centering
    
    \scriptsize 
    \renewcommand{\arraystretch}{1.2} 
    \caption{Impact comparison of \tool with different LLMs.}
    \label{tab:impact}
    \begin{tabular}{l c cccc cc}
        \toprule
        \multirow{2}{*}{LLM Model} & Specifications(components) & \multicolumn{4}{c}{Coverage} & \multicolumn{2}{c}{Failures} \\
        \cmidrule(lr){2-2} \cmidrule(lr){3-6} \cmidrule(lr){7-8}
         & Accuracy & SC & BC & RAC & AAC & MAS & Crash \\
        \midrule
        \tool(GPT-4.1\cite{openai2025gpt41}) & 61/64 & 94.2 & 81.7 & 96.9 & 91.4(100) & 56 & 5 \\
        \tool(Gemini-3-Pro\cite{google2025gemini}) & 63/64 & 94.7 & 84.7 & 90.6 & 81.6(98.8) & 67 & 5 \\
        \bottomrule
    \end{tabular}
    \vspace{10pt}
\end{table}

Given that the LLM serves as the cognitive core of \tool-driving both the semantic coverage analysis and the test oracle, it is crucial to evaluate the system's sensitivity to the underlying model's capabilities. To this end, we conducted a comparative study by replacing the default GPT-4.1 backend with Gemini-3-Pro, while keeping all other algorithmic configurations invariant. Specifically, we re-evaluated the performance of \tool with respect to our three primary research questions: specification conformity, coverage improvement, and failure detection capability. The comparative results are detailed in Table~\ref{tab:impact}.

Regarding specification conformity, the Gemini-backed implementation demonstrated exceptional precision, achieving a Cohen's Kappa coefficient of 1.0. Specifically, only a single instance among the 64 generated components of 15 specifications was identified as erroneous. This performance closely matches the standards observed with the GPT-4.1 backend, indicating that \tool exhibits low sensitivity to model variations and maintains robust reliability across different high-capability LLMs.

In terms of coverage accumulation, we observed consistent growth patterns regardless of the backend model. The intra-agent behavior coverage reached 94.71\%, while the inter-agent path coverage saturated at 83.75\%. Crucially, the comparative analysis suggests that \tool is largely model-agnostic in its exploration capabilities. Despite replacing GPT-4.1 with Gemini-3.0 Pro, the system maintained a comparable level of coverage efficiency. While minor fluctuations were observed in specific sub-metrics, the primary coverage indicators remained stable (and even exhibited slight improvements of 0.54\% and 2.08\%), demonstrating that our method is robust and not heavily reliant on the specific priors of a single foundation model.

The failure categories identified by \tool remain consistent across different underlying LLMs, confirming the stability of our taxonomy. However, \tool configured with Gemini-3 Pro-preview detects a slightly higher number of MAS-specific failures compared to the GPT-4.1 counterpart. We attribute this to the superior reasoning and comprehension capabilities of the Gemini model, which enable a more precise analysis of subtle discrepancies between MAS behaviors and the oracle specifications. Despite this advantage, the overall performance gap between the two models remains narrow, indicating that \tool is robust across capable foundation models.

\section{Threats of Validity}

\textbf{Internal threats to validity.} The primary threat concerns the correctness of LLM-generated specifications. To mitigate hallucinations, we validated our approach via a double-blind human study, where 61 out of 64 components of specifications were strictly consistent with ground truth, yielding a \emph{Linear Weighted Cohen's Kappa} of 0.8512. These results confirm high reliability despite inherent subjectivity. Additionally, the non-deterministic nature of LLMs introduces slight inherent variability in execution traces and coverage across runs.

\textbf{External threats to validity.} External threats mainly relate to generalizability. Our benchmark of 16 open-source applications may not fully reflect the complexity of industrial-scale systems. Second, while our methodology is transferable, the current implementation targets AutoGen; adapting to other frameworks (e.g., LangChain) requires tailoring domain knowledge and prompts. Finally, results may vary when utilizing different LLM backends with distinct reasoning capabilities.
\section{Related Work}
\subsection{Multi-Agent LLM Systems}
Recent advances in LLM have drawn significant attention to MAS in AI~\cite{DBLP:journals/access/CurtoZ25,DBLP:journals/tosem/HeTL25,DBLP:conf/acl/Shahroz0YF025,DBLP:conf/acl/YueZLWWCQ25,DBLP:conf/iclr/HongZCZCWZWYLZR24}. This interest stems from shifting benchmark tasks and a strong demand for autonomous capabilities~\cite{DBLP:conf/nips/LiHIKG23,lyzr_state_ai_agents_h1_2025}. By coordinating specialized agents, MAS can perform stepwise reasoning and tackle complex, multi-stage problems~\cite{DBLP:conf/iclr/ZhangYLYWWCY025}.

Existing MAS have been explored across domains such as drug discovery~\cite{DBLP:journals/corr/abs-2507-09023,DBLP:journals/corr/abs-2507-17852}, financial trading~\cite{DBLP:conf/www/0010ZX0X25,DBLP:conf/kdd/ZhangZXSSQLZ0CZ24}, software engineering~\cite{DBLP:journals/csur/LambiaseCPF25,DBLP:journals/corr/abs-2409-02977}, and society simulation~\cite{DBLP:conf/uist/ParkOCMLB23,DBLP:journals/access/AltunCMEF25}. However, their high degrees of freedom, complexity, and stochastic outputs make them prone to defects in coordination, task execution, inter-agent interaction, and transparency, sometimes with severe consequences~\cite{pan2025why,DBLP:journals/inffus/SapkotaRK26}. Prior work has acknowledged these issues and proposed risk-evaluation benchmarks~\cite{DBLP:journals/corr/abs-2506-04133,zhang2024agent}. They provide only coarse estimates of potential risk and lack testing methods that effectively identify concrete defects. 

To our best knowledge, \tool is the first to propose automated testing solutions for LLM-based multi-agent systems.

\subsection{Testing AI Systems}
Prior testing research on AI systems focuses on neural network accuracy, fairness, and safety, as well as defect analyses of ML frameworks and models~\cite{DBLP:journals/jocnet/SantosMSESSF24,DBLP:journals/remotesensing/TupekZSB23,DBLP:conf/icsm/Cao0HGHS023}. These efforts are typically tailored to specific tasks constrained by fixed output formats. In particular, Keeper~\cite{DBLP:conf/icse/WanLXLHM022} employs invariance-based checks to assess correctness in supervised domains such as image classification. However, since it targets machine learning tasks with categorical outputs, it cannot be extended to MAS, whose outputs are open-ended and require higher-level semantic validation.

Several studies evaluate LLMs and multi-agent systems to estimate the risks of systems~\cite{DBLP:journals/corr/abs-2506-04133,zhan2024injecagent,DBLP:conf/icml/MazeikaPYZ0MSLB24,DBLP:conf/naacl/DongZYSQ24}. Specifically, InfoSynth~\cite{garg2026infosynth} introduces an information-theoretic framework to automatically synthesize diverse and novel benchmarks for assessing the robust reasoning capabilities of LLMs. In a similar vein, TRiSM assesses real-world workflows via a component-coordination score that quantifies inter-agent collaboration quality and tool-use efficiency~\cite{DBLP:journals/corr/abs-2506-04133}. However, such evaluations primarily capture general performance capabilities or quality attributes, but do not directly uncover specific functional defects in MAS.

A recent work~\cite{pan2025why} conducted an empirical study on MAS defects and failure modes, compiling a large-scale dataset to categorize failures. Specifically, it creates a taxonomy comprising three high-level categories: specification and system design failures, inter-agent misalignment, and task verification and termination. However, it focuses on categorization and does not provide automated testing or fixing solutions.
\section{Conclusion}
Testing MAS is inherently challenging due to their non-deterministic interactions and complex state spaces. We present \tool, an automated coverage-guided testing framework designed to systematically explore agent behaviors and detect failures. \tool automatically synthesizes domain-specific test oracles and drives the testing process via a novel feedback loop guided by MAS-specific coverage metrics and a consensus-based defect checking mechanism. We evaluate \tool with a variety of open-source MAS applications and achieve high structural and interaction coverage. It successfully identifies not only generic runtime crashes but also MAS-specific failures, including agent task execution, tool invocation, agent relationships, and system termination anomalies.
\section{Data Availability}
The \tool source code, benchmark applications, and experimental data are available in the Github\footnote{\url{https://anonymous.4open.science/r/ISSTA-FLARE-TEST2026123}}.

\bibliographystyle{ACM-Reference-Format}
\bibliography{ref}

\end{document}